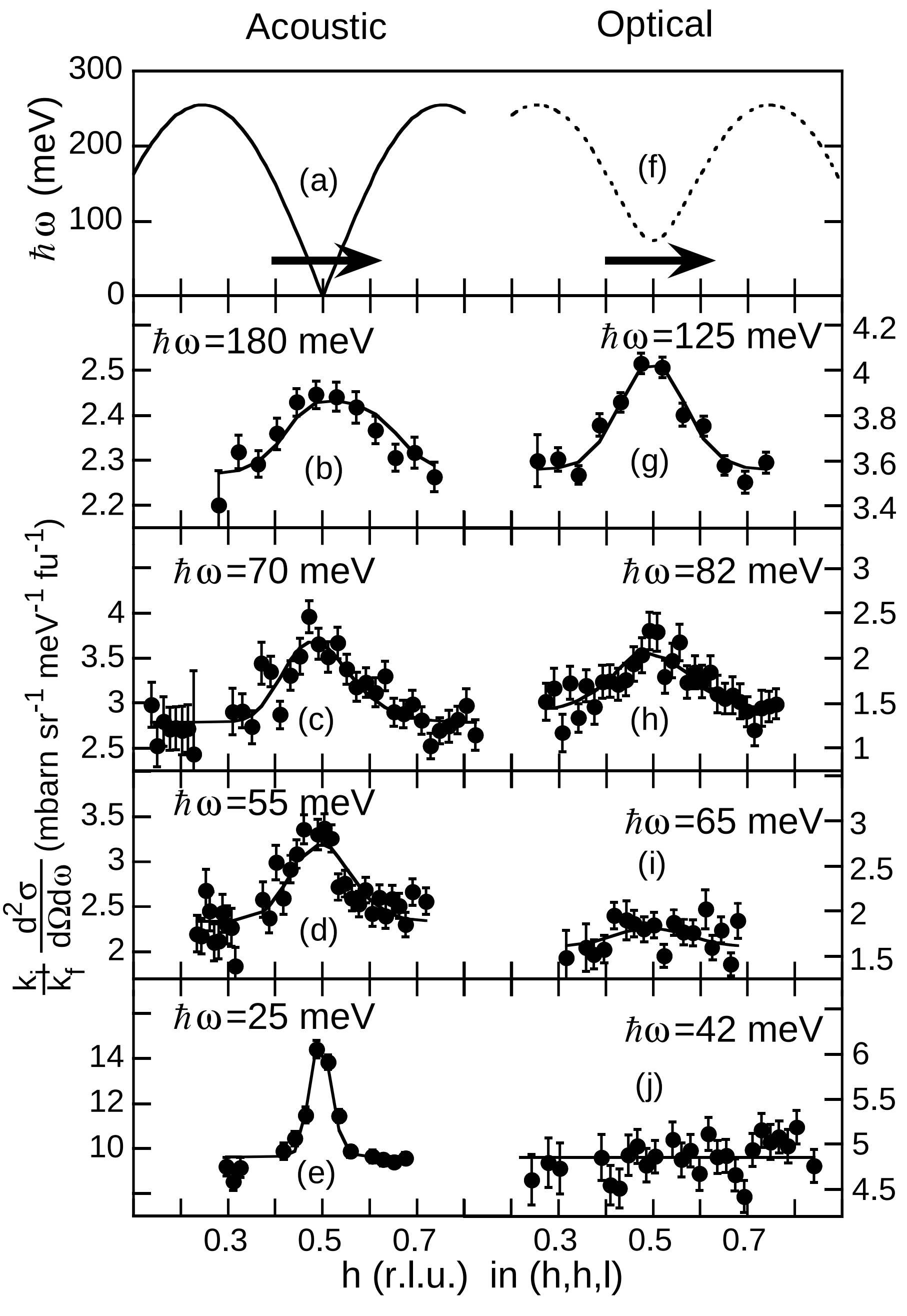

Fig. 1 Hayden et al

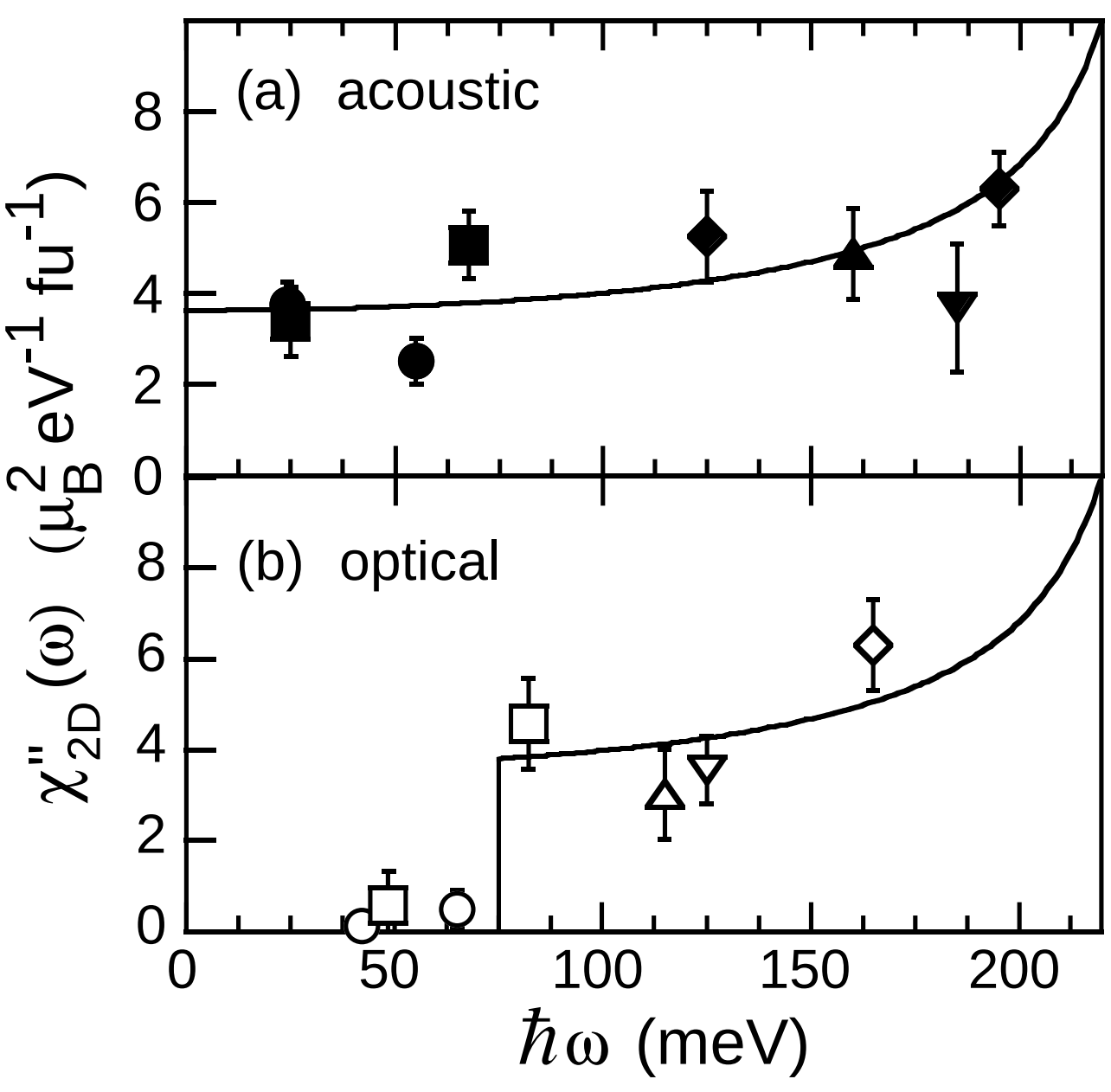

Fig. 2 Hayden et al

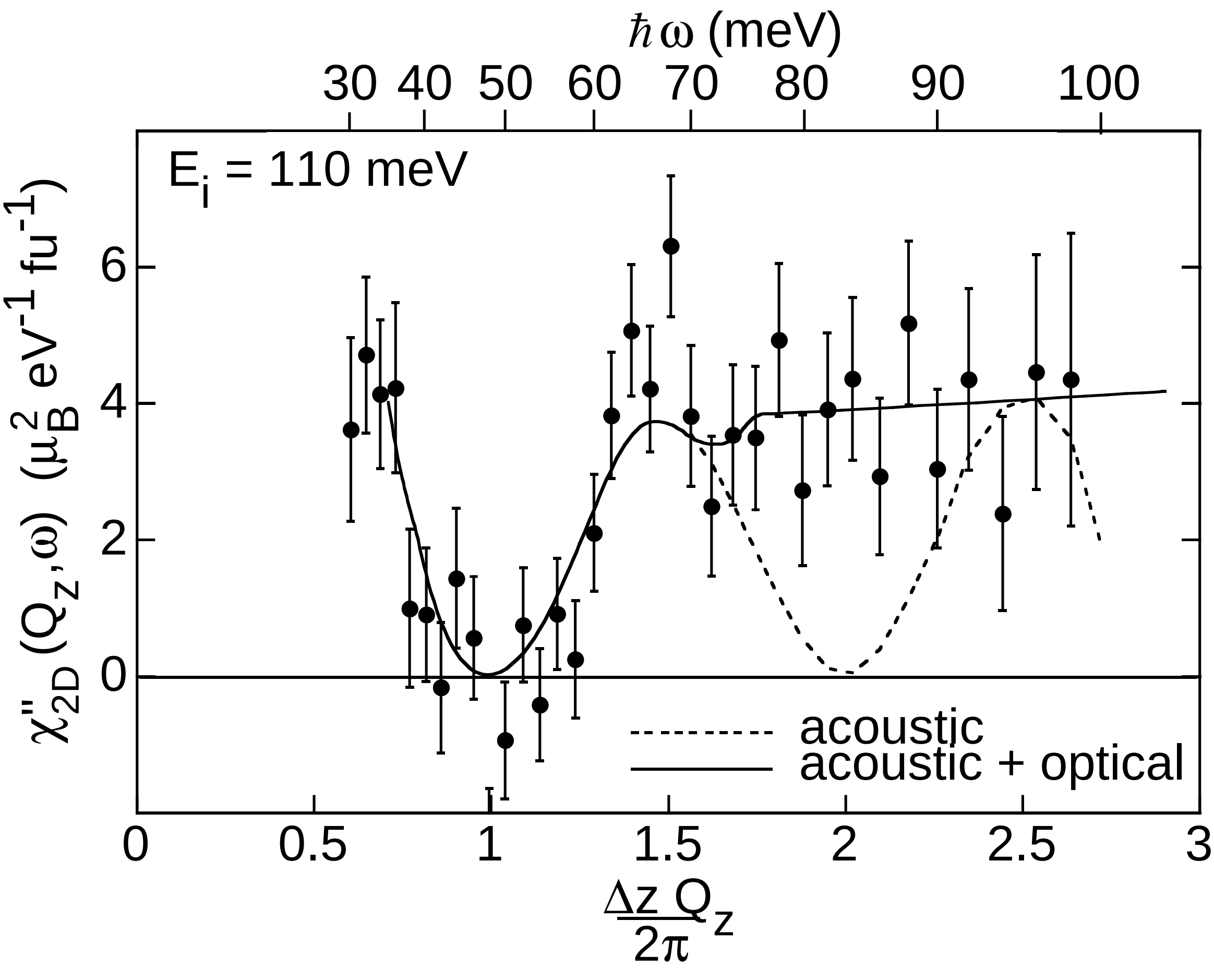

Fig. 3 Hayden et. al.

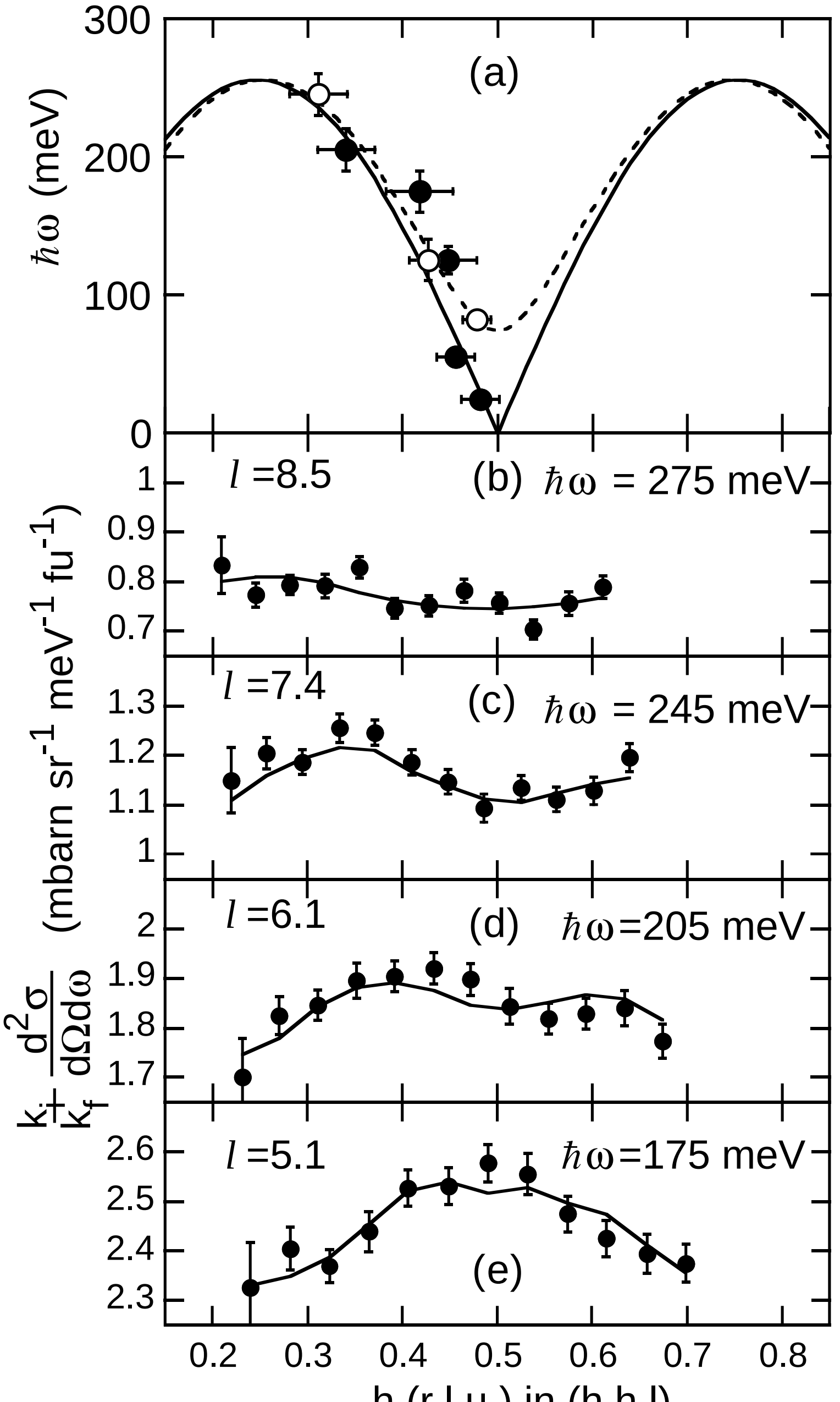

Fig. 4 Hayden et al.

# High-Frequency Spin Waves in $YBa_2Cu_3O_{6.15}$

S. M. Hayden[1], G. Aeppli[2,*], T. G. Perring[3], H. A. Mook[4] and F. Doğan[5]

[1]H. H. Wills Physics Laboratory, University of Bristol, Tyndall Avenue, Bristol BS8 1TL, United Kingdom

[2]AT&T Bell Laboratories, Murray Hill, New Jersey 07974

[3]ISIS Facility, Rutherford Appleton Laboratory, Chilton, Didcot,OX11 0QX, United Kingdom

[4]Oak Ridge National Laboratory, Oak Ridge, Tennessee 37831

[5]Department of Materials Science and Engineering, University of Washington, Seattle, WA 98195

Pulsed neutron spectroscopy is used to make absolute measurements of the dynamic magnetic susceptibility of insulating $YBa_2Cu_3O_{6.15}$. Acoustic and optical modes, derived from in- and out-of-phase oscillation of spins in adjacent $CuO_2$ planes, dominate the spectra and are observed up to 250 meV. The optical modes appear first at $74 \pm 5$ meV. Linear-spin-wave theory gives an excellent description of the data and yields intra- and inter-layer exchange constants of $J_{\|} = 125 \pm 5\,\text{meV}$ and $J_{\perp} = 11 \pm 2\,\text{meV}$ respectively and a spin-wave intensity renormalization $Z_{\chi} = 0.4 \pm 0.1$.

The fundamental building blocks of the high-$T_c$ superconductors are (nearly) square lattices of Cu atoms linked to their nearest Cu neighbours via the p-orbitals of intervening oxygen atoms. The blocks occur in units of one, two or three layers, which, in turn, are separated from each other by spacers serving as charge reservoirs. In general, the superconducting transition temperature rises with the number of $CuO_2$ layers in the units. This means that multilayer materials, most notably $YBa_2Cu_3O_{6+x}$ ($T_c$=93K), have received much more attention than single-layer materials such as $La_{2-x}Sr_xCuO_4$ ($T_c$=38 K). Even so, the magnetic fluctuations, which are popular as a potential source of high-$T_c$ superconductivity, have been characterised over a much wider range of frequencies ($\omega$) and momenta ($\mathbf{Q}$) for $La_{2-x}Sr_xCuO_4$ [1-5] than for $YBa_2Cu_3O_{6+x}$ [6-9] . Indeed, while $La_{2-x}Sr_xCuO_4$, for both insulating and superconducting compositions, has been studied throughout the Brillouin zone for $\hbar\omega$ up to 0.4 eV, magnetic excitations in $YBa_2Cu_3O_{6+x}$ have only been observed for $\hbar\omega$ comparable to, or below, the pairing energy (40 meV) and thus much less than the likely coupling $J_{\|}$ responsible for the antiferromagnetism of the insulating parent. We have consequently performed magnetic neutron scattering experiments to establish the spin dynamics of $YBa_2Cu_3O_{6+x}$ on a high energy scale. The sample chosen for this first measurement is an insulating antiferromagnet with $x = 0.15$. Our study reveals spin-wave excitations up to 0.25 eV. Below $\hbar\omega_g = 74 \pm 5\,\text{meV}$, we observe only acoustic spin waves where pairs of neighbouring spins in adjacent planes rotate in the same sense about their average direction. Above $\hbar\omega_g$, we have discovered optical spin waves [10], where the spins in adjacent planes rotate in opposite directions, thus sensing the restoring force from the inter-planar coupling $J_\perp$. Our data are well described by a linear spin-wave model in which an overall quantum renormalization of the spin-wave intensity $Z_\chi$=$0.4 \pm 0.1$ ( $Z_\chi = 1$ for the classical limit) is included and the intra-planar and inter-planar exchange constants are $J_{\|} = 125 \pm 5\,\text{meV}$ and $J_\perp = 11 \pm 2\,\text{meV}$ respectively.

Experiments were performed on the HET spectrometer at the ISIS pulsed spallation source of the Rutherford-Appleton Laboratory. The experimental techniques for measuring high-frequency spin fluctuations using a direct-geometry chopper spectrometer such as HET are described elsewhere [3]. The sample used in the present investigation was a single crystal of $YBa_2Cu_3O_{6.15}$ with mass 96 g and room temperature lattice constants of a = b = 3.857 Å and c=11.84 Å corresponding to a Néel temperature of 400 K [6]. In common with other large single crystals of $YBa_2Cu_3O_{6+x}$, our sample contains $Y_2BaCuO_5$ as an impurity phase which constitutes up to 15% of the volume. We index reciprocal space using the tetragonal unit cell so that antiferromagnetic Bragg peaks occur at $\left(\frac{1}{2},\frac{1}{2},0\right)$ and related positions, momentum transfers $\left(Q_x,Q_y,Q_z\right)$ in units of $Å^{-1}$ are then at reciprocal space positions $(h,k,l)=\left(Q_x a/2\pi, Q_y b/2\pi, Q_z c/2\pi\right)$. Measurements were made with the $(1\bar{1}0)$ plane coincident with the horizontal scattering plane of the spectrometer. The detectors are 300-mm-long 25-mm-diameter tubes at 2.5 or 4 m from the sample with scattering angles in the range $2.5 < 2\theta < 28$ deg. Absolute unit conversions were performed using a vanadium standard [10]. Data were collected at ambient temperature (T = 296 K).

For the purpose of this experiment, we model $YBa_2Cu_3O_{6+x}$ as a set of weakly coupled $CuO_2$ biplanes, i.e. a square-lattice bilayer antiferromagnet. In this case, the spin dynamics can be described using the Heisenberg Hamiltonian for a single bilayer

$$H=\sum_{ij} J_{\parallel}\,\mathbf{S}_i \bullet \mathbf{S}_j + \sum_{ij'} J_{\perp}\,\mathbf{S}_i \bullet \mathbf{S}_{j'}, \qquad (1)$$

where the first term represents the nearest-neighbour intra-planar coupling between Cu-II spins in the same plane and the second the nearest-neighbour inter-planar coupling between Cu spins in different planes. The introduction of the second term in (1) leads to two branches in the spin-wave dispersion [6] which are labelled according to whether nearest-neighbour spins in different planes rotate together (acoustic mode) or in opposite directions (optical mode) about their average direction. Conventional linear-spin-wave theory of the Holstein-Primakoff type yields the imaginary part of the dynamic susceptibility (per formula unit) for the two modes as:

$$\chi''_{\mathrm{op}}(Q,\omega) = \frac{\pi g^2 \mu_{\mathrm{B}}^2}{\hbar} S \left( \frac{1-\gamma(Q)}{1+\gamma(Q)+J_{\perp}/2J_{\|}} \right)^{\frac{1}{2}} \cos^2\left(\frac{1}{2}\Delta z\, Q_z\right) \delta\left(\omega \pm \omega_{\mathrm{op}}(Q)\right) \qquad (2)$$

and

$$\chi''_{\mathrm{ac}}(Q,\omega) = \frac{\pi g^2 \mu_{\mathrm{B}}^2}{\hbar} S \left( \frac{1-\gamma(Q)+J_{\perp}/2J_{\|}}{1+\gamma(Q)} \right)^{\frac{1}{2}} \sin^2\left(\frac{1}{2}\Delta z\, Q_z\right) \delta\left(\omega \pm \omega_{\mathrm{ac}}(Q)\right), \qquad (3)$$

where the dispersion relations are

$$\hbar\omega(Q) = 2J_{\|}\left(1-\gamma^2(Q) + J_{\perp}/2J_{\|}\left(1 \pm \gamma(Q)\right)\right)^{1/2}$$

(the plus sign is for the acoustic mode), $\gamma(Q) = 1/2\left(\cos\left(aQ_x\right)+\cos\left(aQ_y\right)\right)$ and $\Delta z$( =3.2 Å) is the separation of the $CuO_2$ bilayers. With the coupling terms in Eq. (1), there is no dispersion along the z-direction, only a modulation of the amplitude of the dynamic susceptibility, which may be used to distinguish the two modes. Further, the energy of the optical branch shows a minimum at Q = $\left(\frac{1}{2},\frac{1}{2},l\right)$ of $\hbar\omega_{\mathrm{g}} = 2\sqrt{J_{\|}J_{\perp}}$ .

By varying the incident neutron energy $E_i$, we can arrange to cut across the spin waves at given energy transfers $\hbar\omega$ and values of $Q_z$ ($l$). Figs. 1(b)–(e) show data collected as a function of **Q** parallel to (110) direction for various energy transfers $\hbar\omega$ such that $\sin^2\left(\frac{1}{2}\Delta z\, Q_z\right) \approx 1$ i.e. $l = Q_z c/2\pi = 1.8, 5.5, \ldots$. Under these conditions, we are sensitive only to modes with acoustic character. At all frequencies, we observe a peak near **Q**=$\left(\frac{1}{2},\frac{1}{2},l\right)$ corresponding to the spin waves. Twin peaks, due to counter-propagating spin-wave branches, are not seen due to the poor out-of-plane resolution in the $\left(1\bar{1}0\right)$ direction. Figs. 1(g)-(j) show similar data collected such that $\cos^2\left(\frac{1}{2}\Delta z\, Q_z\right) \approx 1$ i.e. $l = 3.7, 7.3, \ldots$. In this case, we are sensitive only to optical modes. Little or no magnetic scattering is evident below 82 meV. (The area under the 65 meV peak is approximately 10 times smaller than for 82 meV). Our observations are in agreement with thermal neutron scattering studies [6-8], made with $\hbar\omega \leq 42\,\text{meV}$, which observed a $\sin^2\left(\frac{1}{2}\Delta z\, Q_z\right)$ modulation of the magnetic scattering at low energy transfers. In order to make a more quantitative interpretation we use our data to calculate the 2D local- or wavevector-integrated- susceptibility $\chi''_{2\mathrm{D}}(Q_z,\omega) = \int \chi''(Q,\omega)\, d^2Q \Big/ \int d^2Q$ where the Q-integrals are over $(Q_x, Q_y)$ only and $Q_z$ is chosen to be close to an optical or acoustic position. The results are shown in Fig. 2. When the data are plotted in this way, it is clear that for energies $\hbar\omega \geq 82\,\text{meV}$, the optical and acoustic intensities are equal within experimental error. On the other hand, for $\hbar\omega \leq 65\,\text{meV}$, the integrated response at the optical position is consistent with zero. Thus $\hbar\omega_{\mathrm{g}}$ lies between 65 and 82 meV.

A more accurate estimate of the optical gap can be made from Fig. 3. This spectrum shows data collected as a function of $\hbar\omega$ with $E_i$ = 110 meV and integrated over wavevectors with $0.45 \leq h \leq 0.55$, i.e. over the spin-wave cones emanating from $\left(\frac{1}{2}, \frac{1}{2}, l\right)$. Due to energy and momentum conservation in the scattering process, $Q_z$ varies with $\hbar\omega$ in the way indicated by the horizontal axes of Fig. 3. Thus, Fig. 3 displays the local susceptibilities associated with the acoustic and optical modes, weighted by their $Q_z$-dependent structure factors. At lower frequencies ($\hbar\omega \lesssim J_{\parallel}$), the 2-D local susceptibility associated with the acoustic mode is essentially ω-independent and that associated with the optical mode has a step at $\hbar\omega_{\mathrm{g}}$ (see solid lines in Fig. 2). Consequently, in Fig. 3 we expect to observe a simple sinusoidal modulation with $Q_z$ due to the acoustic mode, $\chi''_{2\mathrm{D}}(Q_z,\omega) \propto \sin^2\left(\frac{1}{2}\Delta z\, Q_z\right)$ for $\omega < \omega_{\mathrm{g}}$. For $\omega \geq \omega_{\mathrm{g}}$, a second sinusoidal modulation of almost equal amplitude and $\frac{\pi}{2}$ out of phase, due to the optical mode, is superposed, yielding a $Q_z$- (and ω-) independent signal. By inspection, we can see that above approximately 70 meV the intensity is constant and below this value the characteristic acoustic modulation is observed. A resolution-corrected fit of Eqs. (2)-(4) to the data yields a value [10] for the optical gap of $\hbar\omega_{\mathrm{g}} = 74 \pm 5$ meV.

The optical mode gap is simply proportional to the geometric mean of $J_{\|}$ and $J_{\perp}$, and so by itself, its measurement establishes neither $J_{\perp}$ nor $J_{\|}$. Indeed, to determine $J_{\|}$ and $J_{\perp}$, it is necessary to determine not only $\hbar\omega_g$, but also the spin-wave dispersion as a function of in-plane momentum. We have therefore collected data at energy transfers sufficiently large to make the dispersion obvious. Fig. 4 shows the results obtained for an incident energy $E_i$ = 600 meV. For $\hbar\omega$=175 ± 15 meV, the scattering, though broadened by the dispersion, is still peaked at $\left(\frac{1}{2},\frac{1}{2},l\right)$. For $\hbar\omega$=245 ± 15 meV, the dispersion causes the scattering to be peaked near $(0.35, 0.35, l)$. Finally, for slightly larger energy transfer $\hbar\omega$ = 275 ± 15 meV, there is a much diminished modulation, consistent with a cut-off of approximately 250 meV for single-magnon scattering. The solid lines in Figs. 4(b)-(e) show resolution-corrected fits of linear spin-wave theory. From fitting each energy individually, we determine the dispersion relations shown in Fig. 4(a). A simultaneous fit including the lower frequency data (see solid lines in Figs. (1)-(3)) yields exchange constants $J_{\|} = 125 \pm 5\,\text{meV}$ and $J_{\perp} = 11 \pm 2\,\text{meV}$.

A range of values for $J_{\|}$ have been inferred from other measurements. Rossat-Mignod and co-workers[7] deconvolved reactor-based measurements of low-frequency spin waves and found a spin-wave velocity, near $\left(\frac{1}{2},\frac{1}{2},l\right)$, of $c$=1000 ± 50 meVÅ, which if we assume the absence of further neighbour interactions (confirmed by the present experiment), implies $J_{\|} \approx$ 200 meV. Shamoto and co-workers [8] performed a similar study and found $c$ = 655 ± 100 meVÅ. This is in agreement with the initial slope of the acoustic dispersion curve we have measured. The only pre-existing information on the magnetic energy scale at short wavelengths is from interpretation of the two-magnon Raman effect [11], which suggests, for a model including only a single near neighbour exchange, $J_{\|} \approx$120 meV. Our value for $J_{\|}$ is substantially below that (0.16 eV) [3] for $La_2CuO_4$. This discrepancy may be due to greater off-stoichiometry in our $YBa_2Cu_3O_{6.15}$ sample as well as its larger lattice constants.

It is well known that in two-dimensional quantum antiferromagnets, the spin-wave energies and the overall intensity of the spin waves show quantum renormalizations $Z_c$ and $Z_\chi$ respectively [12,13] when compared with their classical (large S) values. Neutron scattering cannot directly measure $Z_c$, thus the bare exchange constants must be inferred from our effective or measured coupling constants using theoretical estimates [12] of $Z_c \approx 1.18$. In contrast, we are able to measure $Z_\chi$ experimentally by placing our measurements on an absolute intensity scale. Following this procedure, we obtain a value of $Z_\chi = 0.4 \pm 0.1$, close to that given by a $\frac{1}{S}$ spin-wave expansion [13] and indistinguishable from that given by our measurements [5] on $La_2CuO_4$.

Not as much is known about $J_\perp$ as about $J_\parallel$ and $Z_\chi$. Even so, detailed analysis [14] of the essentially zero-frequency magnetic response measured [15] by nuclear resonance for $Y_2Ba_4Cu_7O_{15}$ suggests $5 < J_\perp < 20$ meV. It is also interesting to ask what happens to $J_\perp$ in the metallic and superconducting samples, especially because one explanation [16] for the celebrated spin gap behaviour of $YBa_2Cu_3O_{6+x}$, which is not so clearly seen in the single layer compounds, requires $J_\perp \neq 0$. Magnetic scattering for all superconducting samples of $YBa_2Cu_3O_{6+x}$ displays the modulation characteristic of acoustic spin-wave modes [7,9]. Because scattering has been observed up to $\hbar\omega \approx 45$ meV $= 0.6\hbar\omega_g(x{=}0.15)$, the bilayer coupling for even the optimally doped superconductor is unlikely to be much smaller than 4 meV or 45 K which is 36% ($=0.6^2$) of $J_\perp(x = 0.15)$.

In summary, we have measured, for the first time, single-magnon excitations in the insulating antiferromagnet $YBa_2Cu_3O_{6.15}$ throughout the Brillouin zone. Our results are well described by a linear spin-wave model for a 2D square bilayer, provided an appropriate overall amplitude renormalization is included. The resulting inter- and intra- planar exchange constants are $J_\parallel = 125 \pm 5\,\text{meV}$ and $J_\perp = 11 \pm 2\,\text{meV}$ respectively.

The support of the UK-EPSRC and the US-DOE at Bristol, ISIS and Oak Ridge is gratefully acknowledged.

**Figure Captions**

FIG. 1. (a) and (f) show the dispersion relations for acoustic and optical spin-wave branches respectively. Remaining panels are energy scans showing magnetic scattering from $YBa_2Cu_3O_{6.15}$ for wavevector transfers along $\mathbf{Q} = (hhl)$. (b)-(e) $l$ is chosen to give scattering from acoustic modes. (g)-(j) $l$ is chosen to give scattering from optical modes. Incident energies and $Q_z$ wavevector components are $E_i$ = 600, 110, 75, 75, 600, 110, 75, 75 meV and $l$=5.3, 5.6, 5.5, 1.9, 3.6, 7.2, 7.1, 3.7 for (b)-(e), (g)-(j) respectively.

FIG. 2. The 2-D local susceptibility as a function of energy transfer obtained by integrating over spin-wave peaks and correcting for the $Cu^{2+}$ magnetic form factor, Bose factor and instrumental resolution. (a) acoustic positions and (b) optical positions. The figure is a compilation based on the analysis of data such as those in Fig. 1 with various incident energies.

FIG. 3. Data collected with $E_i$ = 110 meV, integrated over the spin-wave cones in the $(Q_x, Q_y)$ plane near $Q = (\frac{1}{2}, \frac{1}{2}, l)$ and corrected for the $Cu^{2+}$ magnetic form factor. The resulting spectrum is collected along a trajectory in $(Q_z, \hbar\omega)$ defined on the upper and lower axes. Under these conditions the sinusoidal intensity modulation with $Q_z$ will disappear at $\hbar\omega_g \approx 74$ meV. Solid line corresponds to the best description in terms of spin wave theory. Dotted line is the acoustic mode contribution.

FIG. 4. (a) The dispersion relation obtained from independent fits at each energy transfer (all data in paper). Closed circles and solid line are for acoustic modes, open circles and dotted line are optical modes. (b)-(e) Constant energy scans showing high-frequency magnetic scattering from $YBa_2Cu_3O_{6.15}$. Data were collected with $\mathbf{k}_i$ || (001) and $E_i$ = 600 meV.   Counting time was 29 h at 170 μA proton current with a Ta target.  Solid lines are resolution-corrected fits of a linear spin-wave model for a bilayer (see text).

**References**

*Present address: NEC Research Institute, 4 Independence Way, Princeton, NJ 08540